\documentclass[conference]{IEEEtran}
\IEEEoverridecommandlockouts
\usepackage{cite}
\usepackage{amsmath}
\usepackage{amssymb}
\usepackage{amsfonts}
\usepackage{algorithmic}
\usepackage{graphicx}
\usepackage{textcomp}
\usepackage{xcolor}
\usepackage{subfig}
\usepackage{hyperref}
\usepackage[switch]{lineno}
\def\BibTeX{{\rm B\kern-.05em{\sc i\kern-.025em b}\kern-.08em T\kern-.1667em\lower.7ex\hbox{E}\kern-.125emX}}

\begin{document}
\title{Multi-Agent Design Assistant for the Simulation of Inertial Fusion Energy}

\author{%
\IEEEauthorblockN{Meir H. Shachar,
M. Giselle Fernández-Godino\IEEEauthorrefmark{1},
Harshitha Menon\IEEEauthorrefmark{1},
Dane M. Sterbentz\IEEEauthorrefmark{1},\\[0.2em]
Jonathan L. Belof\IEEEauthorrefmark{4},
Yue Hao,
Kevin A. Korner,
Robert N. Rieben,\\[0.2em]
Daniel A. White,
Charles F. Jekel\IEEEauthorrefmark{2},
William J. Schill\IEEEauthorrefmark{2},
}
% \\[0.2em]
\vspace{0.2em}
\IEEEauthorblockA{Lawrence Livermore National Laboratory Livermore, CA 94550, USA 
\\ [0.2em]
% \\[0.8em]
\{shachar1, fernandez48, gopalakrishn1, sterbentz1, belof1, hao1, korner1, rieben1, white37, jekel1, schill1\}@llnl.gov
\\[1em]
}
\thanks{\IEEEauthorrefmark{1} Equal contribution: Fernandez, Menon, Sterbentz.}
\thanks{\IEEEauthorrefmark{2} Principal Investigator.} 
\thanks{\IEEEauthorrefmark{4} Current address: Advanced Micro Devices Santa Clara, CA 95054, USA.}

\IEEEauthorblockN{Ismael D. Boureima}
\vspace{0.2em}
\IEEEauthorblockA{Los Alamos National Laboratory Los Alamos, NM 87545, USA
\\[0.2em]
iboureima@lanl.gov
\\[1em]}

\IEEEauthorblockN{Nathan K. Brown}
\vspace{0.2em}
\IEEEauthorblockA{Sandia National Laboratories Albuquerque, NM 87185, USA
\\[0.2em]
nkbrown@sandia.gov
\\[1em]}
}

\maketitle

\begin{abstract}
Inertial fusion energy promises nearly unlimited, clean power if it can be achieved. However, the design and engineering of fusion systems requires controlling and manipulating matter at extreme energies and timescales; the shock physics and radiation transport governing the physical behavior under these conditions are complex requiring the development, calibration, and use of predictive multiphysics codes to navigate the highly nonlinear and multi-faceted design landscape. We hypothesize that artificial intelligence reasoning models can be combined with physics codes and emulators to autonomously design fusion fuel capsules. In this article, we construct a multi-agent system where natural language is utilized to explore the complex physics regimes around fusion energy. The agentic system is capable of executing a high-order multiphysics inertial fusion computational code. We demonstrate the capacity of the multi-agent design assistant to translate natural-language design goals into simulation execution, physics-emulator construction, analysis, and iterative refinement of capsule-geometry parameters, ultimately achieving simulated ignition within the computational design workflow.
\end{abstract}


\section{INTRODUCTION}
Energy and energy density are the key drivers of civilizational development and wealth creation\cite{smil2017energy}. 
% CJ:: double emdash is not my style, i'd consider to break this into a couple sentEnces, unless you were really trying to put the spoken emphasis on IFE here
Inertial fusion energy (IFE) — the use of high-power lasers or magnetic drive to dynamically confine matter to induce fusion reactions — promises near-infinite, clean energy and a major jump in achievable energy density \cite{abu2024achievement,kritcher2024design, dickman1987nations, pease1958controlled, tollefson2023nuclear, tollefson2024how}. 
The extreme scales needed to achieve these conditions, however, lead to grand challenges in understanding and designing IFE systems.
The challenges arise because both the science of material behavior under extreme conditions and the design of the IFE systems involve many unknowns. A lack of knowledge of the physics induces uncertainty in our designs and lack of knowledge in the sensitivity of our designs creates uncertainty in our ability to adequately constrain material behavior under these conditions.
This significant amount of both aleatoric and epistemic uncertainties is a barrier to progress in IFE development.
%

\begin{figure}[htbp]
\centering
\includegraphics[width=0.45\textwidth]{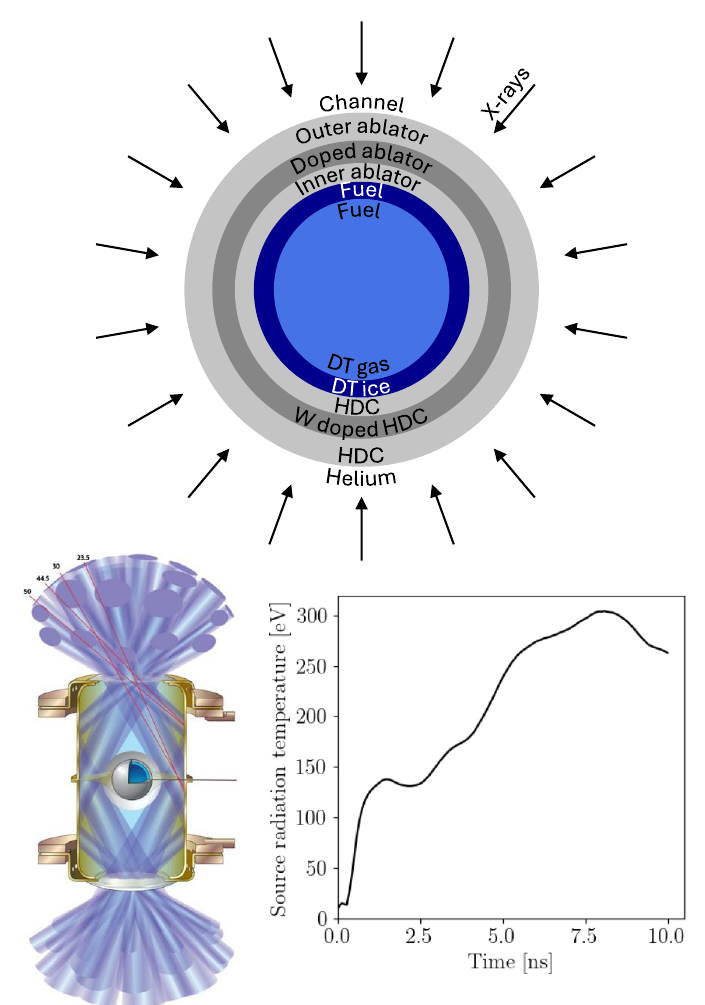}
\caption{Depiction of an inertial fusion energy experiment fielded at the NIF.  The laser impingement onto the hohlraum creates a radiation field that drives the implosion via ablation, with the imposed laser drive a key input to both simulations and experiments of this kind.}
\label{fig1:subfig-a}
\end{figure}

% New paragraph
The principal tools for advancing this field are world-class fusion research facilities — e.g. the National Ignition Facility (NIF) at Lawrence Livermore National Laboratory — and radiation hydrodynamics multiphysics codes running on exascale-class supercomputers. 
In Fig.~\ref{fig1:subfig-a}, we show a schematic of a NIF inertial confinement fusion (ICF) capsule (reproduced, in part, from \cite{moses2008national}). 
An x-ray drive (generated via laser deposition into a hohlraum) delivers energy into the outer capsule layer known as the ablator; this section rapidly ablates launching spherically convergent shockwaves inwards.
These travel through the ablator until they hit the fusion fuel which consists of a solid DT ice shell (which contains the vast majority of the fusion fuel). Inside the DT ice layer, there is a DT gas sphere which, driven by the converging shockwave, collapses to form a hot spot.
As the hotspot is formed, the surrounding DT ice layer is rapidly compressed to high densities. The hotspot ignites the fusion fuel propagating thermonuclear burn consuming the fuel and releasing significantly more energy than the initial laser drive.

We take measurements at extreme conditions to improve the modelling and then use these computational tools to push the boundary of new experiments. 
Such iteration has provided a pathway that advances scientific knowledge, and ultimately led to fusion ignition being demonstrated on Earth~\cite{PhysRevLett.129.075001.LessAuthors}.
Further progress is partially limited by the rate at which we can use modeling and simulation to reason, understand, and prioritize the physical regimes that should be explored next.
When designing the next ICF capsule to try, we point out that in most traditional formulations of optimization the objective function is thought of as a given; in design of NIF capsules, as with many practical engineering and physics challenges, the picture is perhaps less clear cut. 
Using total energy generated as an objective may yield an optimization 
that is not robust to small changes in inputs; looking at general products of $\rho R\times T$ is perhaps a better approach. 
However, there are many reasonable products that one could write down.  
In practical terms, the human designer almost always needs to consider multiple different plausible objective functions and holistically consider their implications.

% New paragraph
% CJ:: I'm going to start to rewrite this as a lit review on ai/ml techniques for physics exploration. Followed by a new paragraph on LLM lit review.
Methods have been developed to utilize machine learning to aid the exploration of designs via computational simulations~\cite{fernandez2024staged,peterson2024toward,sterbentz2023linear,sterbentz2022design,schill2024suppression,schill2023inference}. 
Earlier work also demonstrated that machine-learning models could identify non-obvious ICF design concepts, including random-forest-based discovery of zonal-flow design ideas\cite{peterson2017zonal}.
More recently, automated Bayesian optimization and meta-Bayesian optimization have been applied to direct-drive fusion experiments on OMEGA~\cite{gopalaswamy2025automated,ejaz2024optimizing}.
%
% CJ:: Im going to break out the Professor paper here, since it comes up later in this work. Feel free to expand the impact of some of these other papers.
Generative surrogate models have also been proposed as visualization tools for exploring how design changes affect hydrodynamic simulations in real time~\cite{jekel2024machine}. 
The tool is effectively a fast emulator (inference time of around two microseconds) of the actual multiphysics code (run time of around five minutes). It is built from an ensemble of parameterized multiphysics calculations, and then a generative model is trained to predict the full-field solutions from the parametric inputs to the simulations.
Machine learning techniques for emulating physical processes are widespread and played a role in the recent breakthrough in realizing ignition on NIF~\cite{humbird2019transfer,doi:10.1126/science.adm8201}; such models provide a fast-running surrogate for high-fidelity physics simulation codes. Ultimately, these models in turn can be efficiently used for design optimization, uncertainty quantification, and sensitivity analysis~\cite{nakhleh2021exploring,fernandez2021identifying}.

% New paragraph on LLM lit review
% CJ:: I keep this original citation list, but maybe we should clean it up?
%\cite{lecun2015deep,vaswani2017attention,zaremba2014recurrent,vinyals2015pointer,he2015deep,he2016identity,amodei2015deep,graves2014neural,blog2015unreasonable,hinton1993keeping,krizhevsky2012imagenet,vinyals2015order,huang2019gpipe,yu2015multi,gilmer2017neural,bahdanau2014neural,kaplan2020scaling,brooks2024video,chollet2019measure,legg2008machine,hutter2005universal,lecun2022path, lin2017does} (see also related ideas in relational reasoning \cite{santoro2017simple,santoro2018relational}, image generation \cite{rombach2021high,zhang2023adding,podell2023sdxl} notions and quantification of complexity writ large \cite{aaronson2014quantifying,grunwald2004tutorial}, representation learning \cite{chen2016variational}, program synthesis \cite{austin2021program, li2024guiding}, neural approximation of the solution of partial differential equations \cite{pfaff2020learning,osti_1834708}, and diffusion models \cite{rombach2022high})
Recently, there has been an explosion of research in artificial intelligence (AI) techniques centered around Large Language Models (LLM) \cite{zheng2025learning, dreyer2025mechanistic, doerig2025high, lecun2015deep, vaswani2017attention, zaremba2014recurrent, vinyals2015pointer, he2015deep, he2016identity, amodei2015deep, graves2014neural, blog2015unreasonable, hinton1993keeping, krizhevsky2012imagenet, vinyals2015order, huang2019gpipe, yu2015multi, gilmer2017neural, bahdanau2014neural, kaplan2020scaling, brooks2024video, chollet2019measure, legg2008machine, hutter2005universal, lecun2022path,  lin2017does} (see also related ideas in relational reasoning \cite{santoro2017simple,santoro2018relational}, image generation \cite{rombach2021high,zhang2023adding,podell2023sdxl}, agentic AI and tool use \cite{zheng2025large, bran2024augmenting, dealmeida2025multimodal, miret2025enabling, boiko2023autonomous, kienle2025querycad, zhang2024comprehensive, Hou2025autofea, makatura2023large}, the quantification of complexity \cite{aaronson2014quantifying,grunwald2004tutorial}, representation learning \cite{chen2016variational}, program synthesis \cite{austin2021program, li2024guiding}, neural approximation of the solution of partial differential equations \cite{pfaff2020learning,osti_1834708}, and diffusion models \cite{rombach2022high}). 
Agentic AI is not restricted to LLMs; more generally, agent systems combine policies, tools, memory, environment interaction, evaluation protocols, and scaling considerations~\cite{kim2025towards,zhu2026establishing}.
In the present work, we use an LLM-based reasoning policy to coordinate domain-specific tools and agents for simulation-based ICF design.
LLMs are autoregressive models that predict the next token given a sequence of preceding tokens, which may be drawn from a wide range of contexts.
Each token represents some number of characters, representing partial or complete words, numbers and symbols, and make up the vocabulary of the model. 
LLMs can be aligned to follow human instructions, as demonstrated by \cite{ouyang2022training}; in practice, this is well thought of as a prompt-response pair.
Such models have shown large growth in both raw performance and test-time-compute-based reasoning \cite{jaech2024openai}; this provides a well-founded basis for AI agents to autonomously carry out computational work and perform tasks.

% new paragraph
In this article, we develop a multi-agent design assistant (MADA) built on frontier reasoning models for setting up, running, and analyzing multiphysics simulations of inertial confinement fusion.
MADA is distinct from prior ML-based design studies in its emphasis on agentic orchestration of the simulation-design workflow, including natural-language task specification, simulation execution, emulator construction, analysis, and iterative design recommendation.
Importantly, MADA dynamically builds full-field physics emulators from ensembles of high-fidelity simulations it executes autonomously.
These physics emulators can be immensely useful on their own, allowing users and AI to explore the design space quickly, investigate sensitivities, and to quickly optimize.
Scientific images and plots generated by the emulator are then interpreted by the agents in a multi-modal manner.
The physics emulator constitutes — in our view — a form of \textit{memory} for the physical behavior that can be observed from the simulations; MADA can then drive the full-field emulation and perform reasoning to improve a fusion fuel capsule design within its own design iteration loop, autonomously.
The resulting feedback loop of full-field physics emulators to LLM-based agents and back again leads to fascinating, intelligent exploration and convergence to high-performing and robust  designs. We envision MADA as a first step on the path toward AI-based control systems that will be necessary for the successful design and operation of IFE power plants in the future.

%new paragraph
The present work focuses on simulation-space design, where MADA reasons over the outputs of radiation-hydrodynamics simulations and physics emulators. Translating such workflows to experimental design on facilities such as NIF or OMEGA will require additional steps, including calibration to experimental data, uncertainty quantification, model-discrepancy treatment, and validation against experimental observables. Transfer-learning approaches that adapt models trained on simulations using experimental data provide one possible route for incorporating such information into future MADA workflows~\cite{humbird2022transfer,spears2025predicting}.

\section{MADA ARCHITECTURE AND METHODS}
\subsection{Multiagent architecture and simulation workflow}
We conceive of MADA as a conversation, overseen by a planning agent, between an inverse design agent (IDA), a job management agent (JMA), a simulation agent, and an ML-based physics emulation (which we call Professor) surrogate agent. This multi-agent architecture creates a more flexible (due to the modular nature, allowing swapping out of agents while limiting the effect on the behavior of other agents) and transparent (from being able to review the conversation between agents) framework.

One could conceive of an LLM interface that is completely general; however, we have found it to be practical to limit the arbitrary execution of each AI agent in terms of building a reliable system. Rather than relying on unbounded code generation, we provide each AI agent with a limited set of tools it is allowed to execute. We thus structure MADA as an LLM accepting an arbitrary natural language prompt and responding with the appropriate agent via either natural language or precise execution of the available tools. MADA then serves as both a multi-modal prompt-response computing interface for multiphysics scientific computing and as an autonomous platform for design of ICF capsules.

\begin{figure*}[htbp]
\centering
\includegraphics[width=0.8\textwidth]{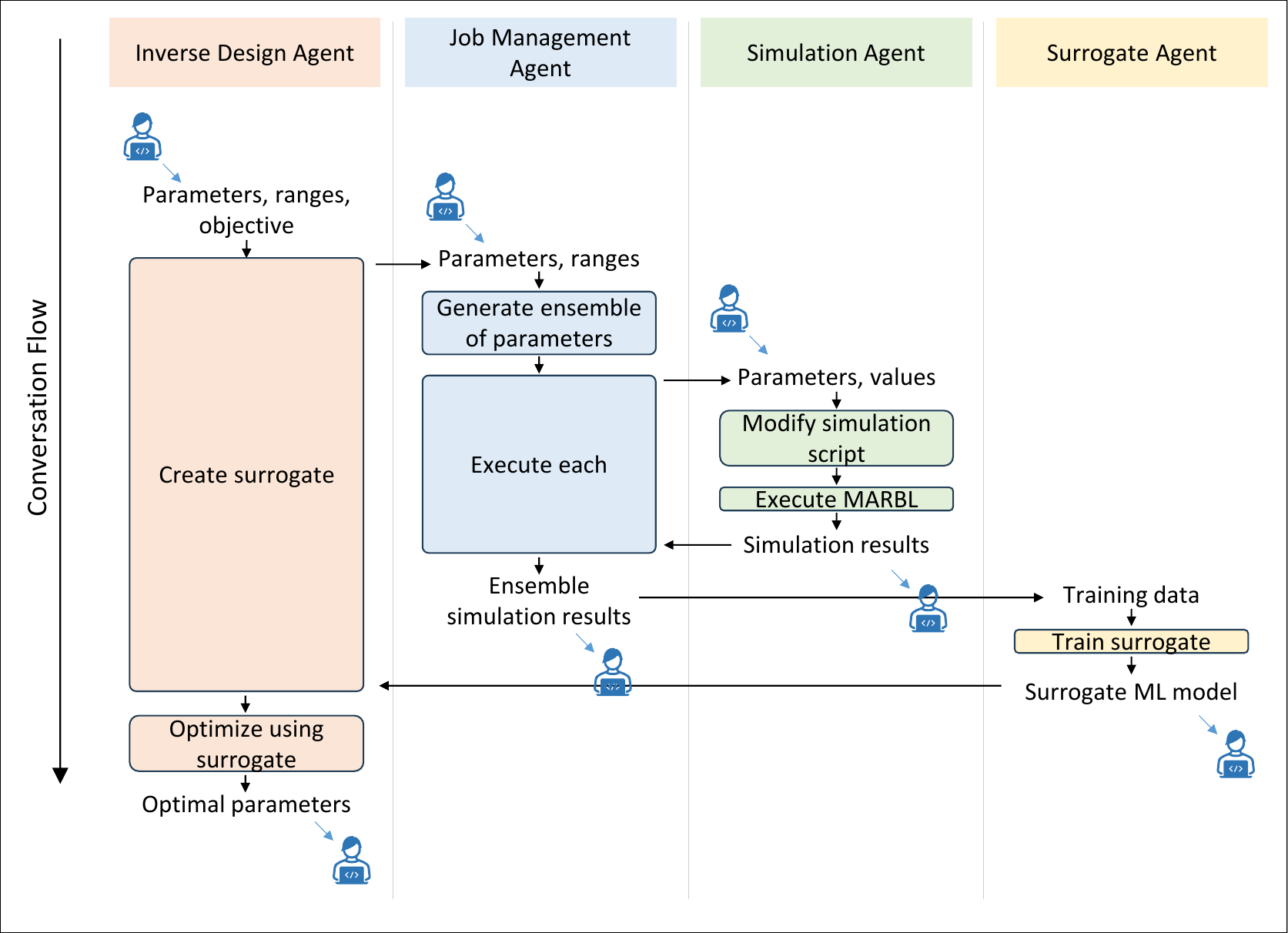}
\caption{The overall architecture and design patterns for the Multi-Agent Design Assistant (MADA) purposed with generating machine learning hydrodynamic surrogates. MADA acts both as an aid to the human ICF designer within an interactive environment and as an autonomous researcher capable of action under broad direction.}
\label{fig1:subfig-b}
\end{figure*}

The task of designing a \textit{good} NIF capsule consists of selecting the sizes and materials of the different layers in the ablator, the x-ray drive profile, and overall size and distribution of the fusion fuel.
To simulate these NIF capsule implosions, we use the high order multiphysics code MARBL \cite{dobrev2012high,dobrev2011curvilinear,anderson2018high,anderson2015monotonicity,vargas2025multi}.

We show in Fig.~\ref{fig1:subfig-b} the architecture and central workflow of MADA. 
The planning agent, which often communicates with the other agents at the insertion points depicted by the blue icon, takes in instructions from the user and passes information to the other agents. 
The IDA is responsible for optimizing a design using a natural language prompt. 
The simulation agent takes in a natural language prompt and modifies inputs to the simulation deck (in MARBL's case a sequence of Lua-language inputs for the multiphysics code) and, possibly, modifying the simulation deck directly.
Having a dedicated IDA agent increases modularity, flexibility, and transparency of inverse design tasks.
The JMA orchestrates the running and management of a suite of HPC job submissions and effectively drives the supercomputer. 
Using an agent to handle job management can offer some unique advantages over traditional job management frameworks, such as monitoring, diagnosing, and restarting failed jobs.
The IDA also identifies parameter ranges specified by the user over which MADA should sample. %Harshitha/Meir clarify this.
Once the simulation is run, the JMA can provide images and reports in the prompt-response feed; alternatively, if prompted to run a large number of simulations to explore a parameter range, the JMA sets up and runs an ensemble of simulations. %Harshitha/Brian mention Merlin here?
A representative MADA design loop makes this multi-agent interaction concrete. The planning agent receives a high-level user objective and decomposes it into tasks for the domain agents. The simulation agent modifies the MARBL input deck, while the JMA submits and monitors the corresponding HPC jobs. After the simulations complete, the IDA analyzes the outputs, identifies parameter ranges or candidate refinements, and may call Professor to train or query a physics emulator. The resulting emulator outputs, plots, and scalar quantities of interest are returned to the agent context, where they can be used to recommend the next design iteration. Concrete prompt-response traces of this workflow, including batch sampling, refinement planning, and convergence toward the burn region, are provided in the Appendix~\ref{sec:appendix_prompts}.
The workflow is flexible in how the planning agent can prompt the other agents. The blue icons demonstrate interaction points relevant to user queries and responses. 
The planning agent will often interact with the other agents through these interaction points, either prompting them or receiving outputs from them.
During a conversation with a user, the planning agent may use several of these pathways in a session.

\subsection{Physics-emulator construction and agent interaction}
Upon completion of the simulations, the IDA may be prompted to either perform additional simulations or train a full-field physics emulator on the field-valued outputs of the code.
In the later case, the LLM calls a PyTorch-based training tool (which we refer to as Professor) to learn the mappings from simulation inputs to the full-field computational solutions \cite{jekel2024machine}. 
The Professor model is trained to emulate the multiphysics code producing full spatial-temporal field variables including density, position, velocity, energy, and temperature.
The Professor tool does this using a deep convolutional generative adversarial network (DCGAN) surrogate model. The DCGAN is trained using data produced by an ensemble of hydrocode simulations. 
We use $512 \times 512$ $r$-$t$ image fields of the density $\rho$, pressure $P$, and other fields from each simulation to train the model. 
Additionally, time-dependent data for the average gas temperature $T$ and areal density $\rho R$ are taken from a .ultra file from simulation, re-discretized to have 512 evenly spaced points (in $T$ vs $\rho R$ space), and then expanded out to create a $512 \times 512$ field image (where all rows are the same). 
These images for $T$ and $\rho R$ are then learned by the GAN along with the $r$-$t$ field images. Combining these into a matrix of size $n \times 512 \times 512$ ($n$ is the number of unique fields) per simulation, we use this to train a GAN surrogate model that correlates these fields to input design parameters (i.e., the thicknesses of various layers of the ICF capsule). 
Typical training for a five-parameter set of interest involves about 3000 simulations for a network size of 100M parameters (see \cite{jekel2024machine} for architectural details) and can be reasonably trained on four MI300A APUs in an hour.
Once the surrogate model is trained, the IDA may be instructed to directly call an optimizer to determine the best parameter values based on a natural-language description of a desired objective function.
Moreover, and perhaps even more interestingly, the IDA may be instructed to use a tool call to feed images directly from the physics emulator to the LLM itself. 
The AI agent may then use the images to reason and make decisions about how to complete the next iterations of the design cycle. This iterative agent-to-agent collaboration can be continued until the specified design objective is satisfied.
%Section Methodology should be double checked by Meir, Dane and Harshitha since its the core of the paper methodology.

\begin{figure}[htbp]
\centering
\subfloat[Basic query resulting in an ICF single simulation which is performed by the MADA system and returned to the user interactively.]{\includegraphics[width=0.40\textwidth]{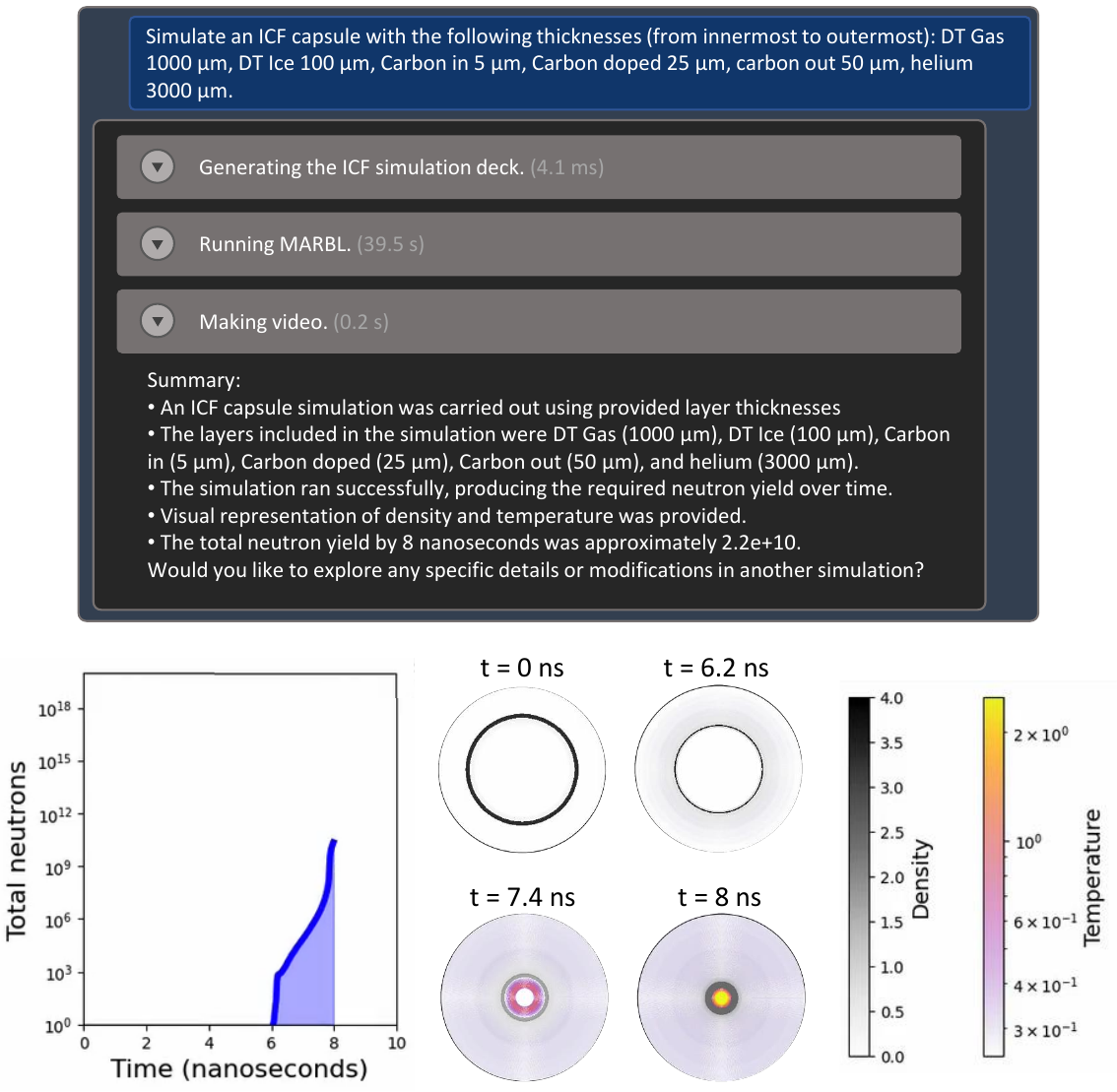}}\label{fig2:subfig-a}
\hfill
\subfloat[Increased complexity query showing the agentic system responding to a user request to submit many jobs sampling within $ 5\%$ of the default value and then generating a plot of temperature vs. areal density paths for the 10 simulations on the right.]{\includegraphics[width=0.40\textwidth]{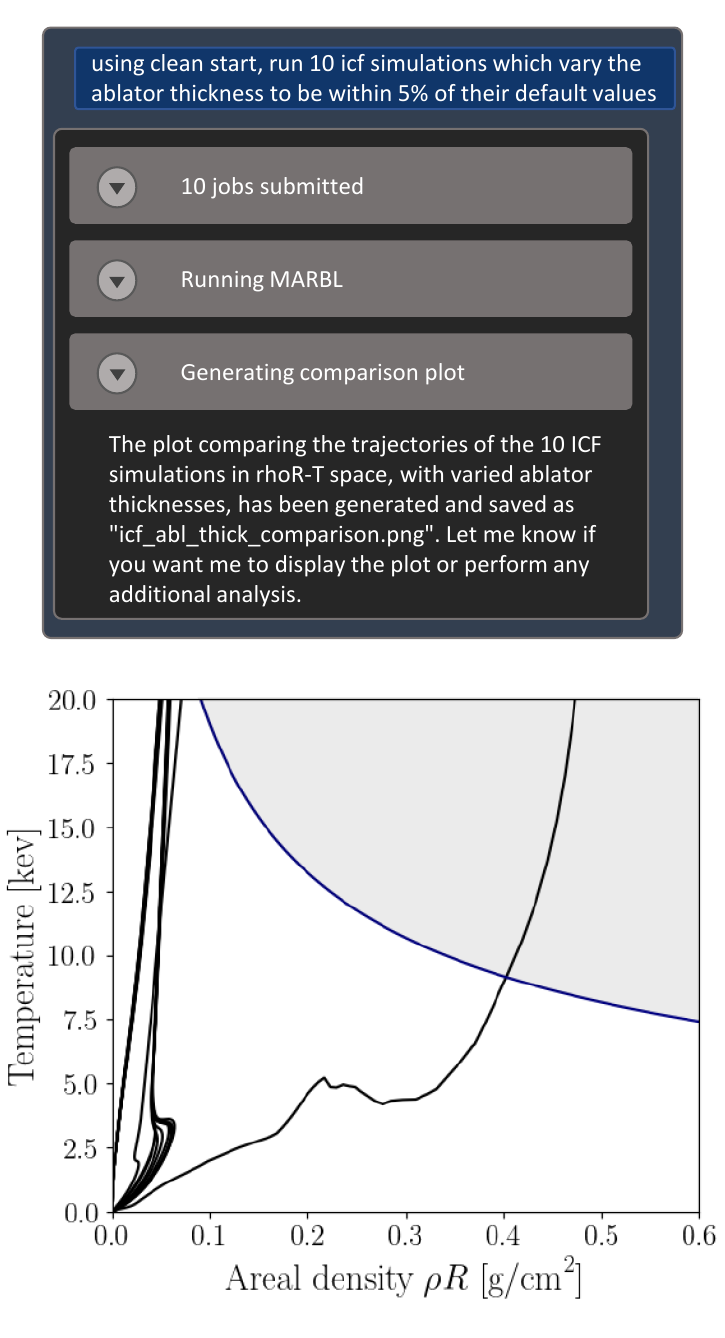}}\label{fig2:subfig-b}
\caption{ Prompt-response pairs for several example queries to the agent system.}
\label{fig2} 
\end{figure}
\section{DEMONSTRATIONS AND DESIGN RESULTS}
\subsection{ Interactive and ensemble simulation execution}
In Fig. \ref{fig2}, we show several illustrative sample prompt-response pairs and an illustration of the MADA interface.
In the first example, the user asks for a particular set of parameters and the planning agent uses the JMA to immediately launch a job interactively and stream the simulation results to the interface.
Both simulation results are kept in the context of the chat interface, which allows the AI system to make comparisons, recommendations, and observations given these high fidelity simulation results.
In the second example, the user asks for Latin hypercube sampling over several parameters and specifies interesting ranges. 
MADA comprehends the instructions and autonomously launches a suite of jobs.

\begin{figure}[htbp]
\centering
\includegraphics[width=0.45\textwidth]{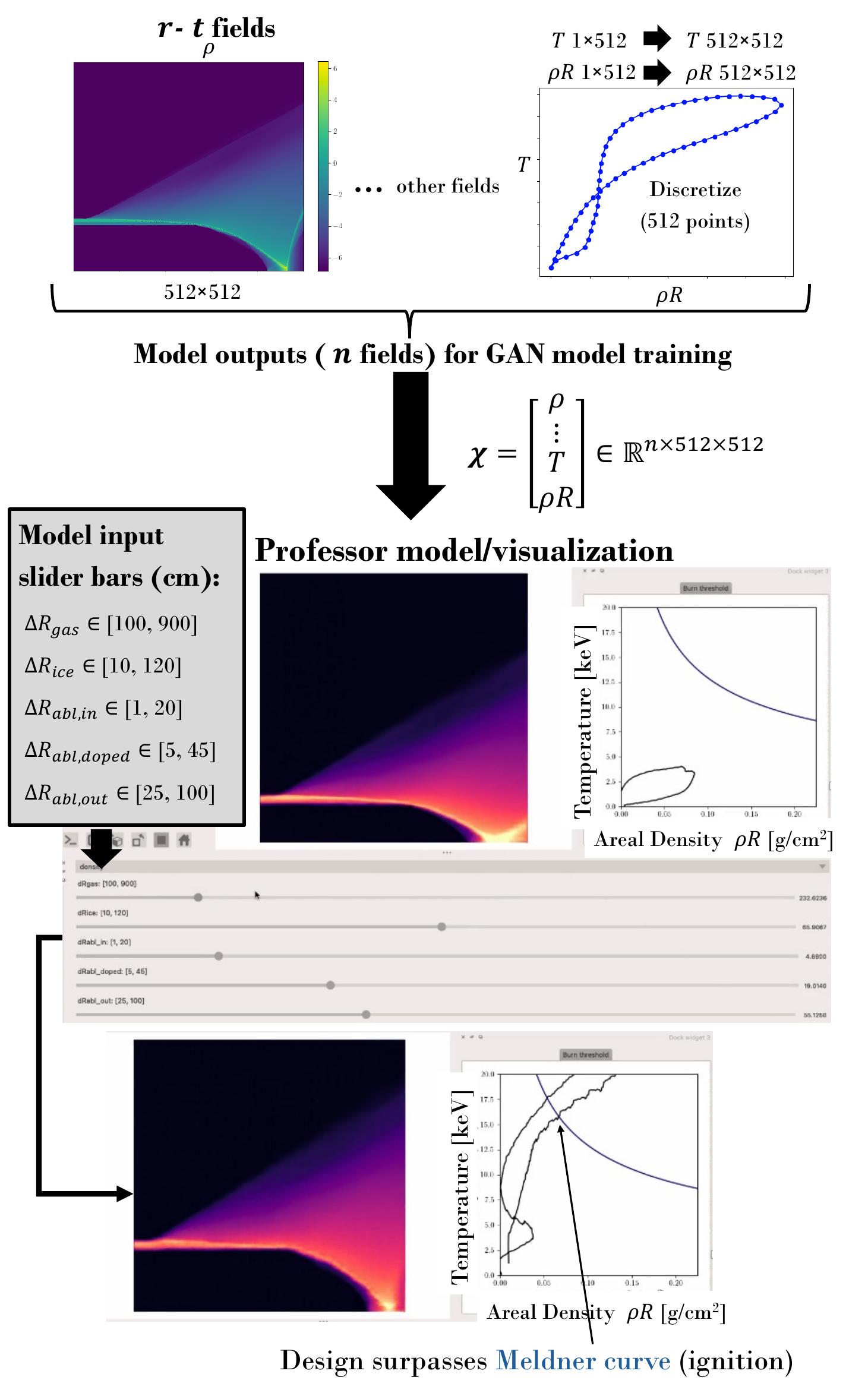}
\caption{Surrogate model produced by Professor, an ML-based multiphysics emulator. Professor is trained to predict full-field data (i.e. $r$-$t$ plots of density, pressure, etc.) and time varying scalars (average fuel temperature and areal density) from input design parameters (e.g. layer thicknesses). The model was trained using results from 3000 simulations. The predictions are summarized using $r$-$t$ plots and state-space plots (fuel temperature vs areal density).}
\label{fig3}
\end{figure}

\subsection{Physics-emulator-guided capsule optimization}
In Fig.~\ref{fig3}, we show some results retrieved from the Professor tool, a full-field, machine learning (ML)-based emulator.
The Professor visualization slider bar interface shows two key outputs from the full-field emulator. The first panel (on the left) is an $r$-$t$ plot of the density, although this can be altered through the interface to show pressure, temperature, or other fields. 
The second panel (on the right) shows a plot of the average gas temperature $T$ versus the areal density $\rho R$. The Meldner curve (navy blue line) is plotted on the $T$ versus $\rho R$ plot to show the threshold above which fusion ignition is achieved. 
The Meldner criterion (mathematically equivalent to the Larsen criteria \cite{tipton2015generalized}) is the locus of points that delineates the boundary between the thermodynamic phase space wherein the DT will achieve runaway burn and where it will not.

The trained DCGAN physics emulator, on its own, can be useful to a user of MADA. By varying the input design parameters using the Professor visualization tool, it is possible to produce a design that surpasses the Meldner curve and achieves ignition (i.e., where the black path exceeds the Meldner curve).
The corresponding $T$--$\rho R$ trajectory for a particular configuration can be quickly produced from the physics emulator and compared against the Meldner curve.
The user may use this interactively to see how far particular combinations of parameters lead to large excursions across this threshold — and hence would be likely to ignite and produce a gain in energy.
Alternatively, the user may prompt the AI agent to utilize this with an optimizer or by directly ingesting image outputs from the physics emulator to aid in a reasoning-based approach to improve the design.
This enables MADA to provide highly flexible optimization by specifying the objective at runtime; or, even being less precise, and specify high-level goals in English and letting the agent try out different objectives.
This flexibility can be immensely useful in ICF design, where the human designer needs to consider different objective functions and holistically account for their implications.
We use a visual-feedback-based approach, in which Professor results from the emulator, as an image resembling those in Fig.~\ref{fig3}, are supplied to the model for any given parameters, and the model uses those to reason about which designs to try next.

% \subsection{AI optimized ICF target using physics emulator}

We now turn to the study of a flexible optimization strategy provided by MADA. We specify high-level goals in natural language and let the AI agent try out different objectives.

\begin{figure}[htbp]
\centering
\includegraphics[width=0.45\textwidth]{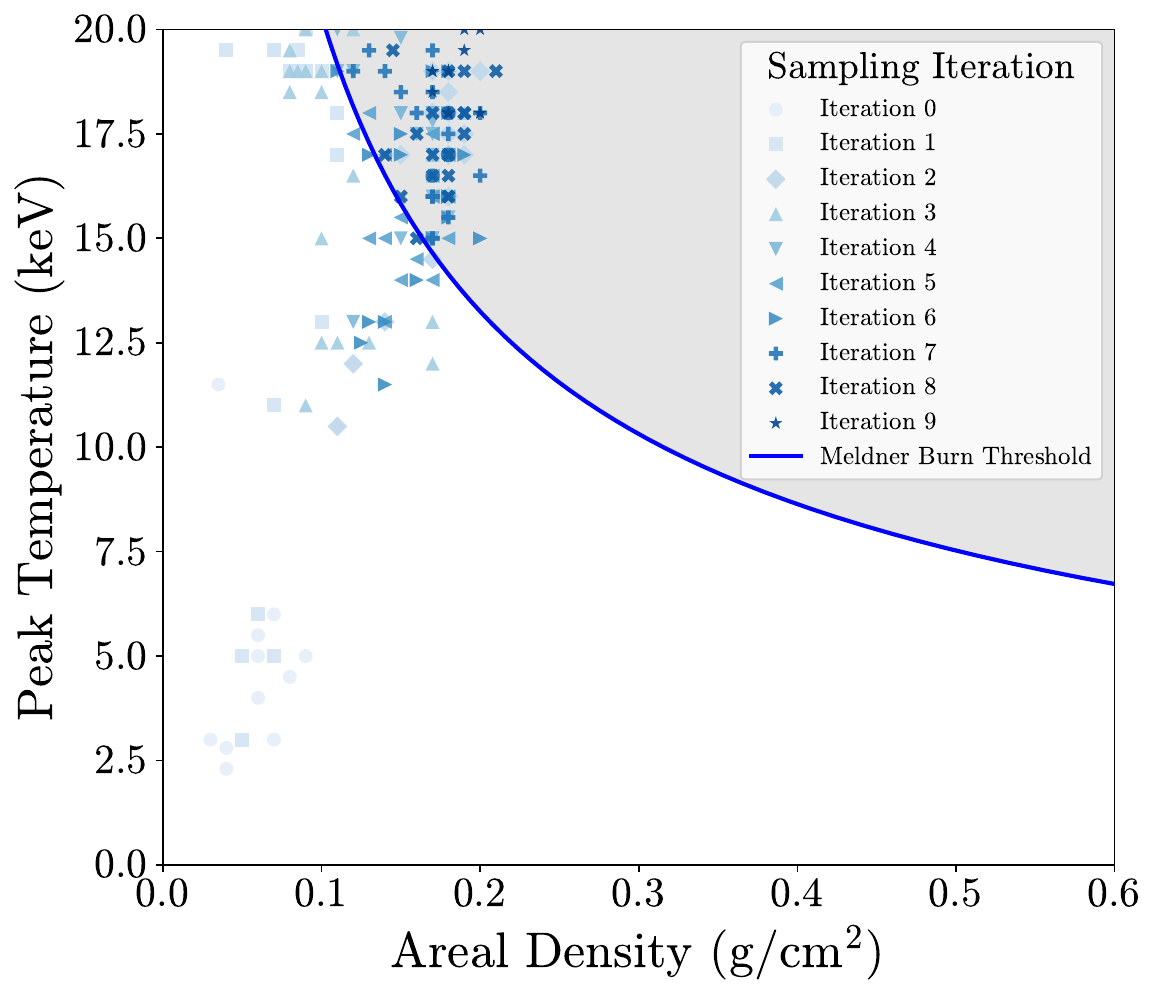}
\caption{Agent-guided iterative sampling converging toward the burn region. In each iteration, the AI agent samples 20 new parameter configurations, shown as colored markers by iteration. Early iterations explore broadly across the design space, while later iterations increasingly focus on high-performing regions. The blue curve represents the Meldner burn threshold; samples lying above this curve indicate entry into the burn region. The progression illustrates the AI agent's growing sensitivity to key parameters and its ability to steer sampling toward fusion-relevant regimes.}
\label{fig4:subfig-a}
\end{figure}

\begin{figure}[htbp]
\centering
\subfloat[]{\includegraphics[width=0.40\textwidth]{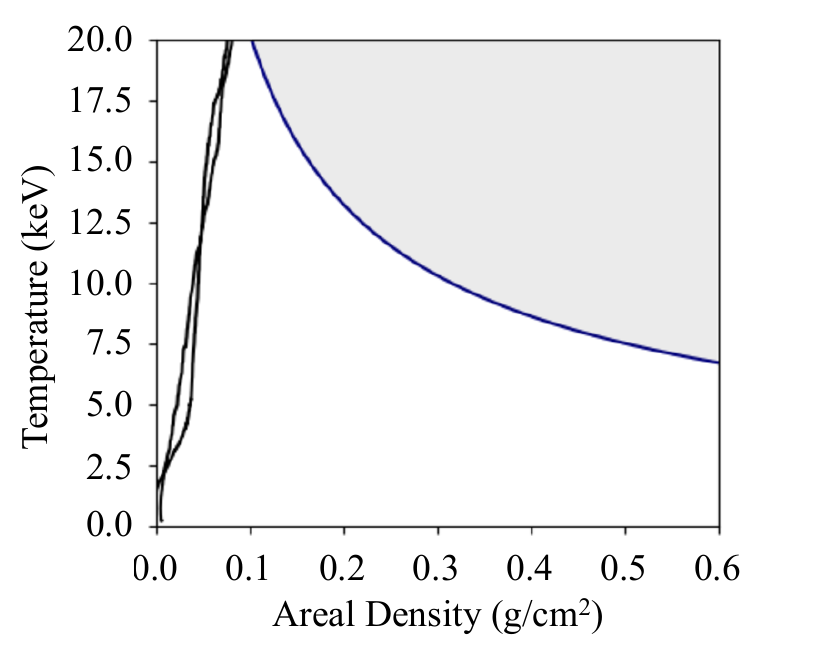}\label{fig4:subfig-b}}
\hfill
\subfloat[]{\includegraphics[width=0.40\textwidth]{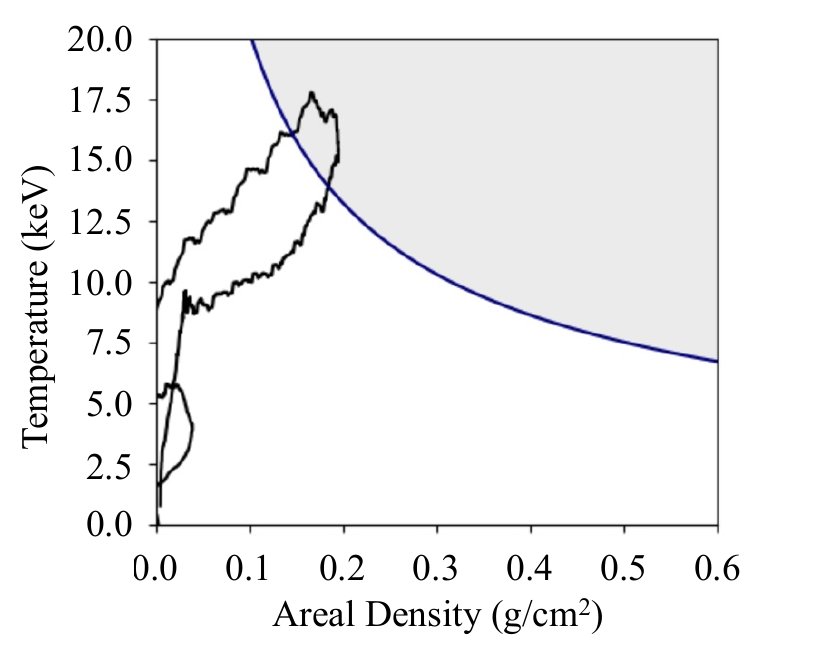}\label{fig4:subfig-c}}
\hfill
\subfloat[]{\includegraphics[width=0.40\textwidth]{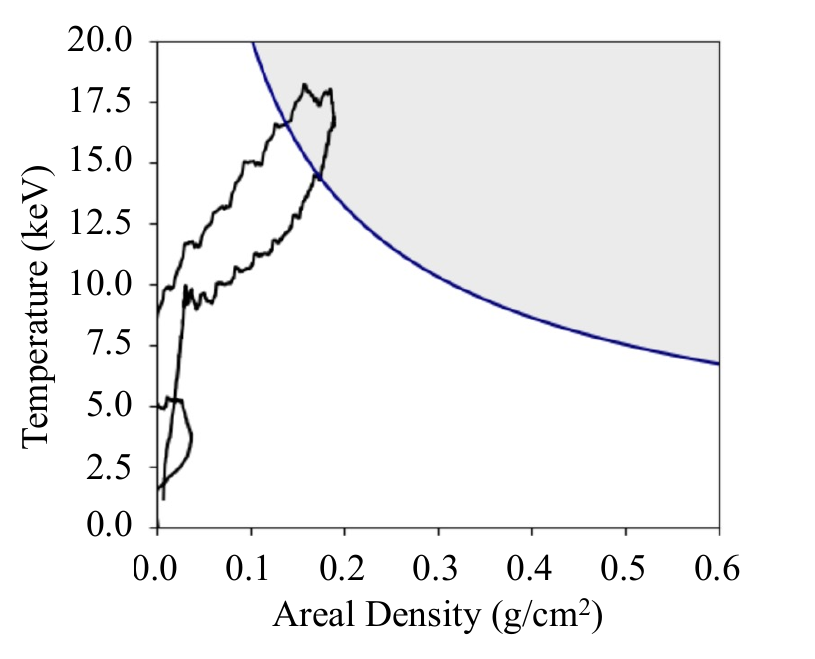}\label{fig4:subfig-d}}
\caption{The figure shows in blue the Meldner burn threshold and in black the $T$–$\rho R$ trajectory for a given parameter configuration. Panel (a) shows the trajectory corresponding to the best-performing sample identified in the initial iteration, panel (b) shows the best sample in the final iteration, and panel (c) shows the plot corresponding to the globally best parameter configuration.}
\label{fig4}
\end{figure}
In Figs.~\ref{fig4:subfig-a} and \ref{fig4}, we study the usage of the professor physics design tool by MADA. 
We show in Fig.~\ref{fig4:subfig-a}, a sequence of nine iterations taken by MADA. In each iteration, MADA uses a trained Professor emulator to try a small handful of different values of parameters and then reasons about what is good or bad about that set of choices and how to improve things. Evidently, as the iterations progress, the samples from each iteration move up and to the right, towards denser, hotter fuel. 
By iteration 9, all samples are above the ignition threshold and MADA finds a global best performer. We show several examples taken from the best performers of early versus late iterations; clearly, the $T$--$\rho R$ trajectories of the later iterations exceed the Meldner criterion resulting in dramatic improvements from a design that would fail to burn, Fig.~\ref{fig4:subfig-b}, to one, Fig.~\ref{fig4:subfig-c}, that robustly burns and is a near neighbor of the global optima Fig.~\ref{fig4:subfig-d}.
Interestingly, this visual reasoning-based feedback approach succeeds in improving the capsule design.
Though a more traditional optimization approach would also be successful here, the visual-feedback approach is compelling in terms of explainability, flexibility, and the ability to guide the optimization (e.g. by conversing with subject-matter experts or by providing the AI agent with the latest findings in literature).

\section{CONCLUSIONS AND FUTURE WORK}

In conclusion, we have shown the design and overall behavior of an LLM-based reasoning agent for performing multiphysics calculations for IFE.
The model is very flexible admitting a wide range of both interactive and autonomous behavior with different time horizons that is steered through natural language.
The AI agents can do focused tasks across deck generation, parameter modification, interactive simulations, ensemble studies, post-processing, emulator training, optimization, and visual self-feedback.
The last, in particular, led to outstanding performance in terms of semi-autonomous discovery of optimal designs; we think of this as a form of physics-simulation-specific memory available for the model to use.
The ability to offload these tasks to MADA will streamline IFE design workflows while reducing user time and effort. 
In addition, the MADA architecture prioritized modularity, allowing the expertise of different agents to be transferable to other workflows.
It is also possible to use MADA's LLM-based natural-language interfaces to teach new users to the IFE design process.
We close with an as-of-yet unproven conjecture that is suggested by our results.
It has been observed that LLMs are meta optimizers and can perform in context learning using an implicit learned gradient descent algorithm related to their original training data \cite{von2023transformers}.
By arming the LLM with multiphysics simulation codes and fast-running emulators, we have extended this meta-optimization capability to an entirely new domain of multiphysics with no additional training of the language model.
% CJ:: Meta comment, it would be nice to highlight a case where the LLM said do to somehting, but our simulation code says otherwise, thus demonstrating this 
Providing AI systems with the ability to execute the tools scientists use to explore new physics regimes may be the onset of a new frontier of autonomous scientific discovery. 

\section*{ACKNOWLEDGMENTS}

This work was performed under the auspices of the U.S. Department of Energy by Lawrence Livermore National Laboratory under Contract DE-AC52-07NA27344 and was supported by the LLNL-LDRD Program under Project No. 21-SI-006 and Project No. 24-ERD-005. Lawrence Livermore National Security, LLNL-JRNL-2011708. The authors would like to acknowledge the programmatic leadership and support from Teresa Bailey and Rob Neely. The authors at LLNL, LANL and SNL gratefully acknowledge funding from the NNSA Advanced Simulation and Computing program.

\section*{DATA AVAILABILITY}

There are no publicly available research data or software supporting this paper. Requests for further information or data should be sent to the authors.

\bibliographystyle{unsrt}
\bibliography{research}

\vspace{12pt}

\clearpage

%\appendices
\section*{\textbf{APPENDIX}\\ USER PROMPTS AND AGENT RESPONSES}
\label{sec:appendix_prompts}
Here we present representative user prompts and responses generated by the MADA agent during the optimization workflow. These responses highlight the agent's ability to interpret high-level user intent, generate structured experiment plans, and refine its sampling strategy based on observed simulation outcomes. Rather than reproducing the raw interface screenshots, the prompts and result tables are transcribed below. Throughout, layer thicknesses (\texttt{dRgas}, \texttt{dRice}, \texttt{dRabl\_in}, \texttt{dRabl\_doped}, \texttt{dRabl\_out}) are in $\mu$m, peak temperature \texttt{peak\_T} is in keV, and peak areal density \texttt{peak\_$\rho R$} is in g/cm$^2$. A dash (``---'') indicates that the solver failed to report a value for that metric.

\subsection{User prompt}
The user initiates the workflow with the following prompt (a batch optimization task pointing at a YAML configuration file):

\begin{quote}
Run a batch optimization of parameters in the configuration file \url{vis_n210808_ult_5x512x512.yaml}.

\emph{First, show the plan:} List 20 diverse configurations within valid parameter ranges. Explain how you sampled them. Explain how burn-region entry will be evaluated and how results will be recorded. Burn region is above the blue curve and entry is when the black curve ($T$ vs $\rho R$) crosses above the blue Meldner curve. Describe the final results presentation format.

\emph{Wait for my confirmation. Then execute exactly one batch:} Evaluate all 20 configurations sequentially without pausing mid-batch. Output a single results table with columns \texttt{[trial\_id, dRgas, dRice, dRabl\_in, dRabl\_doped, dRabl\_out, peak\_T, peak\_$\rho R$, distance, in\_burn\_region]}.
\end{quote}

\subsection{Initial sampling strategy}
The agent identifies the parameter bounds from the YAML file,
\begin{center}
\begin{tabular}{ll}
\texttt{dRgas} & $[100, 900]$~$\mu$m\\
\texttt{dRice} & $[10, 120]$~$\mu$m\\
\texttt{dRabl\_in} & $[1, 20]$~$\mu$m\\
\texttt{dRabl\_doped} & $[5, 45]$~$\mu$m\\
\texttt{dRabl\_out} & $[25, 100]$~$\mu$m\\
\end{tabular}
\end{center}
and proposes Latin-Hypercube Sampling (LHS) to cover the 5-D space evenly with only 20 points: each variable range is divided into 20 equal-probability intervals, one value is drawn from every interval, and the draws are permuted across variables. This preserves diversity, ensures edge and interior coverage, and avoids the clustering that pure random sampling can create. The resulting 20 candidate configurations are listed in Table~\ref{tab:appendix_lhs}.

\begin{table}[h]
\centering
\caption{Initial 20 candidate configurations proposed by the agent via Latin-Hypercube Sampling. All values lie strictly inside their respective bounds.}
\label{tab:appendix_lhs}
\resizebox{\columnwidth}{!}{%
\begin{tabular}{rrrrrr}
\hline
id & \texttt{dRgas} & \texttt{dRice} & \texttt{dRabl\_in} & \texttt{dRabl\_doped} & \texttt{dRabl\_out}\\
\hline
1  & 100 & 10  & 1  & 5  & 25\\
2  & 142 & 16  & 2  & 8  & 32\\
3  & 216 & 30  & 4  & 12 & 40\\
4  & 274 & 43  & 6  & 17 & 48\\
5  & 358 & 55  & 7  & 20 & 56\\
6  & 416 & 68  & 9  & 24 & 64\\
7  & 500 & 80  & 11 & 28 & 72\\
8  & 558 & 92  & 13 & 31 & 80\\
9  & 642 & 105 & 15 & 35 & 88\\
10 & 700 & 118 & 17 & 39 & 96\\
11 & 784 & 25  & 19 & 43 & 34\\
12 & 842 & 38  & 4  & 10 & 42\\
13 & 900 & 51  & 8  & 14 & 65\\
14 & 316 & 64  & 12 & 18 & 79\\
15 & 474 & 77  & 16 & 22 & 87\\
16 & 632 & 110 & 20 & 26 & 95\\
17 & 210 & 20  & 5  & 30 & 50\\
18 & 370 & 35  & 10 & 34 & 70\\
19 & 530 & 60  & 14 & 38 & 90\\
20 & 790 & 85  & 18 & 42 & 100\\
\hline
\end{tabular}%
}
\end{table}

\subsection{First batch results}
The results of evaluating all 20 initial configurations are given in Table~\ref{tab:appendix_batch0}. The agent summarizes:
\begin{itemize}
\item Only Trial~5 (\texttt{dRgas}=358, \texttt{dRice}=55, \texttt{dRabl\_in}=7, \texttt{dRabl\_doped}=20, \texttt{dRabl\_out}=56) achieved burn-region entry (distance~$=0.0$).
\item The highest reported \texttt{peak\_$\rho R$} among successful simulations was $0.09$ (Trial~4); however, that point did not enter burn.
\item Several higher-$\rho R$ candidates failed to return metrics, possibly indicating solver non-convergence at extreme parameters.
\end{itemize}
The agent proposes (i)~refining around Trial~5's neighborhood to push $\rho R$ higher while maintaining burn-region entry, and (ii)~investigating solver failures in larger-radius cases.

\begin{table*}[t]
\centering
\caption{First batch evaluation results. Trial~5 was the only configuration to enter the burn region.}
\label{tab:appendix_batch0}
\resizebox{\textwidth}{!}{%
\begin{tabular}{rrrrrrrrrl}
\hline
id & \texttt{dRgas} & \texttt{dRice} & \texttt{dRabl\_in} & \texttt{dRabl\_doped} & \texttt{dRabl\_out} & \texttt{peak\_T} & \texttt{peak\_$\rho R$} & distance & \texttt{in\_burn\_region}\\
\hline
1  & 100 & 10  & 1  & 5  & 25  & 2.3  & 0.040 & 12.0 & False\\
2  & 142 & 16  & 2  & 8  & 32  & 2.8  & 0.040 & 9.0  & False\\
3  & 216 & 30  & 4  & 12 & 40  & 4.0  & 0.060 & 1.0  & False\\
4  & 274 & 43  & 6  & 17 & 48  & 5.0  & 0.090 & 7.0  & False\\
5  & 358 & 55  & 7  & 20 & 56  & 19.5 & 0.070 & 0.0  & True\\
6  & 416 & 68  & 9  & 24 & 64  & 6.0  & 0.070 & 3.5  & False\\
7  & 500 & 80  & 11 & 28 & 72  & 3.0  & 0.030 & 10.0 & False\\
8  & 558 & 92  & 13 & 31 & 80  & ---  & ---   & ---  & False\\
9  & 642 & 105 & 15 & 35 & 88  & ---  & ---   & ---  & False\\
10 & 700 & 118 & 17 & 39 & 96  & ---  & ---   & ---  & False\\
11 & 784 & 25  & 19 & 43 & 34  & ---  & ---   & ---  & False\\
12 & 842 & 38  & 4  & 10 & 42  & 5.5  & 0.060 & 8.0  & False\\
13 & 900 & 51  & 8  & 14 & 65  & 11.5 & 0.035 & 1.0  & False\\
14 & 316 & 64  & 12 & 18 & 79  & 3.0  & 0.070 & 12.0 & False\\
15 & 474 & 77  & 16 & 22 & 87  & ---  & ---   & ---  & False\\
16 & 632 & 110 & 20 & 26 & 95  & ---  & ---   & ---  & False\\
17 & 210 & 20  & 5  & 30 & 50  & 4.5  & 0.080 & 8.5  & False\\
18 & 370 & 35  & 10 & 34 & 70  & 5.0  & 0.060 & 8.0  & False\\
19 & 530 & 60  & 14 & 38 & 90  & ---  & ---   & ---  & False\\
20 & 790 & 85  & 18 & 42 & 100 & ---  & ---   & ---  & False\\
\hline
\end{tabular}%
}
\end{table*}

\subsection{Refinement plan}
After analyzing the first batch, the agent ranks the leading trials --- Trial~4 ($\rho R=0.09$, no burn), Trial~5 ($\rho R=0.07$, burn), Trial~17 ($\rho R=0.08$, no burn), and Trials~6 and 14 ($\rho R=0.07$, no burn) --- and identifies the following structure:
\begin{itemize}
\item \emph{Observed clusters.} A ``moderate-radius cluster'' (\texttt{dRgas}~$\approx 300\text{--}420$~$\mu$m, \texttt{dRice}~$\approx 50\text{--}70$~$\mu$m; Trials~4--6 and 17); a ``high-radius cluster'' (\texttt{dRgas}~$\geq 550$~$\mu$m) that frequently produced solver nulls. Only Trial~5 entered burn, so its immediate neighborhood is the prime exploitation area.
\item \emph{Sensitivity indications.} Changing \texttt{dRgas} or \texttt{dRice} moved $\rho R$ strongly; \texttt{dRabl\_in} and \texttt{dRabl\_doped} appeared to modulate temperature but were less sampled; several nulls suggest solver fragility at extreme \texttt{dRgas} and high \texttt{dRabl\_out}.
\end{itemize}
The agent then proposes a refined 20-configuration strategy split into exploitation (small perturbations around Trial~5, Table~\ref{tab:appendix_exploit}), exploration (broader coverage of moderately-sampled zones, Table~\ref{tab:appendix_explore}), and micro-sweep configurations. The realized set of 20 refined configurations, together with their outcomes, is reported in Table~\ref{tab:appendix_batch2}.

\begin{table}[h]
\centering
\caption{Exploitation configurations: $\pm$ small steps around Trial~5 to test whether burn can be entered at higher $\rho R$.}
\label{tab:appendix_exploit}
\resizebox{\columnwidth}{!}{%
\begin{tabular}{rrrrrrl}
\hline
id & \texttt{dRgas} & \texttt{dRice} & \texttt{dRabl\_in} & \texttt{dRabl\_doped} & \texttt{dRabl\_out} & note\\
\hline
E1 & 330 & 55 & 7 & 20 & 56 & lower gas radius\\
E2 & 385 & 55 & 7 & 20 & 56 & higher gas radius\\
E3 & 358 & 45 & 7 & 20 & 56 & lower ice\\
E4 & 358 & 65 & 7 & 20 & 56 & higher ice\\
E5 & 358 & 55 & 5 & 20 & 56 & lower \texttt{Rabl\_in}\\
E6 & 358 & 55 & 9 & 20 & 56 & higher \texttt{Rabl\_in}\\
E7 & 358 & 55 & 7 & 17 & 56 & lower \texttt{Rabl\_doped}\\
E8 & 358 & 55 & 7 & 23 & 56 & higher \texttt{Rabl\_doped}\\
\hline
\end{tabular}%
}
\end{table}

\begin{table}[h]
\centering
\caption{Exploration configurations: broaden coverage in moderately-sampled zones while avoiding extreme regimes that caused solver nulls (first six of the exploration set, as shown by the agent).}
\label{tab:appendix_explore}
\resizebox{\columnwidth}{!}{%
\begin{tabular}{rrrrrr}
\hline
id & \texttt{dRgas} & \texttt{dRice} & \texttt{dRabl\_in} & \texttt{dRabl\_doped} & \texttt{dRabl\_out}\\
\hline
X1 & 300 & 50 & 10 & 22 & 60\\
X2 & 420 & 60 & 6  & 15 & 50\\
X3 & 450 & 70 & 12 & 25 & 60\\
X4 & 380 & 40 & 8  & 18 & 65\\
X5 & 395 & 90 & 10 & 28 & 70\\
X6 & 500 & 55 & 8  & 18 & 55\\
\hline
\end{tabular}%
}
\end{table}

\subsection{Second batch results}
The refined batch (Table~\ref{tab:appendix_batch2}) sharply improves performance. The agent summarizes:
\begin{itemize}
\item Burn-region success rate jumped to $15/20$ trials (75\%).
\item New best areal density: Trial~14 reached $\rho R = 0.17$ in the burn region (\texttt{dRgas}=500, \texttt{dRice}=55).
\item Parameter trends: \texttt{dRice}~$\approx 55$~$\mu$m is robust for burn entry (both $45$ and $65$ retained access with modest $\rho R$ effect); raising \texttt{dRabl\_in} to 9 or \texttt{dRabl\_doped} to 23 lifted $\rho R$ to $0.11$ while maintaining burn; a larger \texttt{dRgas} ($500$~$\mu$m) with moderate ice produced the highest $\rho R$ yet; very high (\texttt{dRice}~$\geq 70$) or very low (\texttt{dRice}~$\leq 30$) ice tended to fail burn or reduce $\rho R$.
\end{itemize}
The agent recommends centering a local search around \texttt{dRgas}~$\approx 500$, \texttt{dRice}~$\approx 50\text{--}60$, \texttt{dRabl\_in}~$8\text{--}10$, \texttt{dRabl\_doped}~$18\text{--}25$, \texttt{dRabl\_out}~$55\text{--}65$, and investigating why modest increases in \texttt{dRabl\_out} ($\geq 70$) suppress burn.

\begin{table*}[t]
\centering
\caption{Second (refined) batch results. Fifteen of twenty configurations entered the burn region.}
\label{tab:appendix_batch2}
\resizebox{\textwidth}{!}{%
\begin{tabular}{rrrrrrrrrl}
\hline
id & \texttt{dRgas} & \texttt{dRice} & \texttt{dRabl\_in} & \texttt{dRabl\_doped} & \texttt{dRabl\_out} & \texttt{peak\_T} & \texttt{peak\_$\rho R$} & distance & \texttt{in\_burn\_region}\\
\hline
1  & 330 & 55 & 7  & 20 & 56 & 13.0 & 0.100 & 0.0  & True\\
2  & 385 & 55 & 7  & 20 & 56 & 19.0 & 0.080 & 0.0  & True\\
3  & 358 & 45 & 7  & 20 & 56 & 19.0 & 0.085 & 0.0  & True\\
4  & 358 & 65 & 7  & 20 & 56 & 19.0 & 0.100 & 0.0  & True\\
5  & 358 & 55 & 5  & 20 & 56 & 20.0 & 0.090 & 0.0  & True\\
6  & 358 & 55 & 9  & 20 & 56 & 17.0 & 0.110 & 0.0  & True\\
7  & 358 & 55 & 7  & 17 & 56 & 19.5 & 0.070 & 0.0  & True\\
8  & 358 & 55 & 7  & 23 & 56 & 18.0 & 0.110 & 0.0  & True\\
9  & 300 & 50 & 10 & 22 & 60 & 5.0  & 0.050 & 7.5  & False\\
10 & 420 & 60 & 6  & 15 & 50 & 19.0 & 0.090 & 0.0  & True\\
11 & 450 & 70 & 12 & 25 & 60 & 5.0  & 0.070 & 12.0 & False\\
12 & 380 & 40 & 8  & 18 & 65 & 19.5 & 0.040 & 0.0  & True\\
13 & 395 & 90 & 10 & 28 & 70 & 3.0  & 0.050 & 6.0  & False\\
14 & 500 & 55 & 8  & 18 & 55 & 18.0 & 0.170 & 0.0  & True\\
15 & 340 & 30 & 6  & 24 & 45 & 11.0 & 0.070 & 1.5  & False\\
16 & 410 & 65 & 9  & 19 & 75 & 6.0  & 0.060 & 7.0  & False\\
17 & 358 & 55 & 7  & 20 & 50 & 19.5 & 0.070 & 0.0  & True\\
18 & 358 & 55 & 7  & 20 & 60 & 19.5 & 0.085 & 0.0  & True\\
19 & 358 & 58 & 7  & 20 & 56 & 19.0 & 0.090 & 0.0  & True\\
20 & 358 & 52 & 7  & 20 & 56 & 19.5 & 0.070 & 0.0  & True\\
\hline
\end{tabular}%
}
\end{table*}

\subsection{Final optimization round}
In the final round (Table~\ref{tab:appendix_batch10}), all 20 configurations crossed the Meldner threshold and entered the burn region. The agent observes:
\begin{itemize}
\item The highest $\rho R$ this round is $0.20$ (Trials~5, 7, 12), matching but not exceeding the record $0.21$ from an earlier round.
\item \texttt{dRabl\_out}~$= 55\text{--}56$~$\mu$m at $(560,68)$ again yields $0.20$; an outer thickness of $61$~$\mu$m recovers $0.20$ but with a slight positive distance ($0.1$).
\item A slightly larger gas radius ($570$~$\mu$m) with $68$~$\mu$m ice also reaches $0.20$, indicating a small plateau around \texttt{dRgas}~$560\text{--}570$~$\mu$m.
\item Variations in inner/doped thickness (\texttt{dRabl\_in}~$7\text{--}9$, \texttt{dRabl\_doped}~$17\text{--}19$) have negligible impact ($\pm 0.01$ in $\rho R$).
\end{itemize}
The agent concludes that the parameter space has plateaued at $\rho R \approx 0.20\text{--}0.21$, with the best overall design being \texttt{dRgas}=560, \texttt{dRice}=68, \texttt{dRabl\_in}=8, \texttt{dRabl\_doped}=18, \texttt{dRabl\_out}=58~$\mu$m ($\rho R = 0.21$), and that further gains likely require expanding the design space beyond the current five thicknesses.

\begin{table*}[t]
\centering
\caption{Final optimization round. All 20 configurations entered the burn region; the design space has plateaued at $\rho R \approx 0.20\text{--}0.21$.}
\label{tab:appendix_batch10}
\resizebox{\textwidth}{!}{%
\begin{tabular}{rrrrrrrrrl}
\hline
\texttt{trial\_id} & \texttt{dRgas} & \texttt{dRice} & \texttt{dRabl\_in} & \texttt{dRabl\_doped} & \texttt{dRabl\_out} & \texttt{peak\_T} & \texttt{peak\_$\rho R$} & distance & \texttt{in\_burn\_region}\\
\hline
1  & 550 & 68 & 8 & 18 & 58 & 20.0 & 0.19 & 0.0 & True\\
2  & 555 & 68 & 8 & 18 & 58 & 18.0 & 0.18 & 0.0 & True\\
3  & 560 & 68 & 8 & 18 & 58 & 18.0 & 0.18 & 0.0 & True\\
4  & 565 & 68 & 8 & 18 & 58 & 18.0 & 0.18 & 0.0 & True\\
5  & 570 & 68 & 8 & 18 & 58 & 18.0 & 0.20 & 0.0 & True\\
6  & 575 & 68 & 8 & 18 & 58 & 18.0 & 0.18 & 0.0 & True\\
7  & 560 & 68 & 8 & 18 & 55 & 20.0 & 0.20 & 0.0 & True\\
8  & 560 & 68 & 8 & 18 & 56 & 19.0 & 0.18 & 0.0 & True\\
9  & 560 & 68 & 8 & 18 & 57 & 19.0 & 0.17 & 0.0 & True\\
10 & 560 & 68 & 8 & 18 & 59 & 18.5 & 0.17 & 0.0 & True\\
11 & 560 & 68 & 8 & 18 & 60 & 18.0 & 0.18 & 0.0 & True\\
12 & 560 & 68 & 8 & 18 & 61 & 18.0 & 0.20 & 0.1 & True\\
13 & 555 & 66 & 8 & 18 & 58 & 19.0 & 0.18 & 0.0 & True\\
14 & 555 & 70 & 8 & 18 & 58 & 19.0 & 0.18 & 0.0 & True\\
15 & 565 & 66 & 8 & 18 & 58 & 18.0 & 0.18 & 0.0 & True\\
16 & 565 & 70 & 8 & 18 & 58 & 19.5 & 0.19 & 0.0 & True\\
17 & 560 & 68 & 7 & 17 & 58 & 19.0 & 0.18 & 0.0 & True\\
18 & 560 & 68 & 7 & 19 & 58 & 18.0 & 0.18 & 0.0 & True\\
19 & 560 & 68 & 9 & 17 & 58 & 19.0 & 0.18 & 0.0 & True\\
20 & 560 & 68 & 9 & 19 & 58 & 19.0 & 0.17 & 0.0 & True\\
\hline
\end{tabular}%
}
\end{table*}

\end{document}